\documentclass[12pt]{article}

\usepackage{epsf}
\usepackage{amsmath}
\usepackage{amssymb}
\usepackage{graphicx}
\usepackage{dcolumn}
\usepackage{bm}

\begin{document}

\def\thefootnote{\fnsymbol{footnote}}

\begin{center}
\Large\bf\boldmath
\vspace*{2.0cm}SuperIso: A program for calculating the\\ 
isospin asymmetry of $B \rightarrow K^* \gamma$ in the MSSM
\unboldmath
\end{center}

\vspace{0.6cm}
\begin{center}
F. Mahmoudi\footnote{Electronic address: \tt nazila.mahmoudi@tsl.uu.se\\
\hspace*{0.48cm} URL: \tt http://www3.tsl.uu.se/$\sim$nazila/}\\[0.4cm]
{\sl High Energy Physics, Uppsala University, Box 535, 75121 Uppsala, Sweden}
\end{center}
\vspace{0.6cm}
\begin{abstract}
\noindent We present a program for calculating the isospin symmetry breaking of the $B \rightarrow K^* \gamma$ decay in the MSSM with minimal flavor violation. This program calculates the NLO supersymmetric contributions to the isospin asymmetry, using the effective Hamiltonian approach and within the QCD factorization method. We show that isospin symmetry breaking proves to be a very restrictive observable, in particular in the mSUGRA parameter space. The program also calculates the inclusive branching ratio associated to $b \rightarrow s \gamma$ transition, as a comparison reference.
\end{abstract}
\vspace{0.3cm}
\section{Introduction} 
\noindent In spite of the numerous phenomenological achievements of the Standard Model, there are still theoretical motivations which impel particle physicists to believe in new physics beyond the standard model. Among these motivations, one can name in particular the hierarchy problem or dark matter/energy problems. As an attempt to solve these problems, supersymmetry (SUSY) appears as a highly motivated and studied model during the past decades.\\
\\
We consider here the Minimal Supersymmetric extention of the Standard Model (MSSM), with minimal flavor violation. 
Since no supersymmetric partner has been discovered, supersymmetry has to be a broken symmetry. A large effort has been done to understand the nature of supersymmetry breaking, and the most successful mechanisms are believed to be the gravity mediated (SUGRA) in which the SUSY breaking sector communicates with the MSSM sector via only gravitational interactions, anomaly mediated (AMSB) in which SUSY breaking sector is transmitted to the observable sector through the super-Weyl anomaly, and gauge mediated (GMSB) which is mediated by gauge interactions. The boundary conditions at high scales in the mentioned scenarios reduce the number of free parameters of the MSSM, allowing feasible phenomenological studies.\\
\\
During the past few years, rare B decays, and in particular those associated to $b \rightarrow s \gamma$ transitions  have been extensively used in order to constrain SUSY parameter space. The observable commonly used for this purpose is the inclusive branching ratio corresponding to this transition. In \cite{wilsoncoeff} we have introduced a new observable, the isospin asymmetry, and showed that this observable can produce complementary information to the branching ratio, and in some regions of the parameter spaces, even more restrictive constraints.\\
\\
We introduce here \verb?SuperIso? as a tool to calculate the isospin symmetry breaking for the three main SUSY breaking scenarios, {\it i.e.} mSUGRA, AMSB and GMSB. The program is modular so that the user can insert its own model (as long as it produces the SUSY mass spectra and couplings according to the SUSY Les Houches Accord conventions). The code allows also to calculate the inclusive branching ratio, so that comparison between the two observables is possible. \\
\\
The paper is organized as follows. The relevent equations for the calculation of the isospin asymmetry are summarized in section \ref{calculation}. In section \ref{content} the \verb?SuperIso? package is described, and the installation and compilation instructions can be found in section \ref{compilation}. In section \ref{sample} we explain how to use \verb?SuperIso?, with input and output samples. Finally some results are presented in section \ref{result}.
%
\vspace{2cm}
\section{Calculation of the isospin symmetry breaking}%
\label{calculation}
\noindent The isospin asymmetry $\Delta_{0-}$ can be written to leading order \cite{kagan}:
\begin{equation}
\Delta_{0-} =\mbox{Re}(b_d-b_u) \;\;,
\end{equation} 
where the spectator-dependent coefficients $b_q$ reads:
\begin{equation}
b_q = \frac{12\pi^2 f_B\,Q_q}{\bar m_b\,T_1^{B\to K^*} a_7^c}\left(\frac{f_{K^*}^\perp}{\bar m_b}\,K_1+ \frac{f_{K^*} m_{K^*}}{6\lambda_B m_B}\,K_{2q} \right)\;\;.
\end{equation}
The functions $K_1$ and $K_{2q}$ can be written in function of the Wilson coefficients $C_i$ at scale $\mu_b$\footnote{Please note that we adopt the conventions of \cite{kagan,bosch}, in which $C^{(0)}_1(M_W) = 1$ and $C^{(0)}_2(M_W) = 0$, therefore $C_1$ and $C_2$ are inverted here compared to \cite{wilsoncoeff}.}:
\begin{eqnarray}
K_1 &=& -\left( C_6(\mu_b) + \frac{C_5(\mu_b)}{N} \right) F_\perp+ \frac{C_F}{N}\,\frac{\alpha_s(\mu_b)}{4\pi}\,\left\{\left( \frac{m_b}{m_B} \right)^2 C_8(\mu_b)\,X_\perp \right.\\
&&\left. -C_1(\mu_b) \left[ \left(\frac43\ln\frac{m_b}{\mu_b} + \frac23 \right) F_\perp - G_\perp(x_{cb})\right] + r_1 \right\}\nonumber \;\;,\\
\nonumber\\
K_{2q} &=& \frac{V_{us}^* V_{ub}}{V_{cs}^* V_{cb}}\left( C_1(\mu_b) + \frac{C_2(\mu_b)}{N} \right) \delta_{qu} + \left( C_4(\mu_b) + \frac{C_3(\mu_b)}{N} \right) \\
&&+ \frac{C_F}{N}\,\frac{\alpha_s(\mu_b)}{4\pi} \left\{ C_1(\mu_b)\left( \frac43\ln\frac{m_b}{\mu_b} + \frac23 - H_\perp(x_{cb}) \right) + r_2 \right\} \;\;,\nonumber
\end{eqnarray}
where $N=3$ and $C_F=4/3$ are color factors, and:
\begin{eqnarray}
r_1 &=& \left[\, \frac83\,C_3(\mu_b) + \frac43\,n_f(C_4(\mu_b)+C_6(\mu_b)) - 8(N C_6(\mu_b)+C_5(\mu_b)) \right] F_\perp \ln\frac{\mu_b}{\mu_0} + \dots \,, \nonumber\\
r_2 &=& \left[ -\frac{44}{3}\,C_3(\mu_b) - \frac43\,n_f(C_4(\mu_b)+C_6(\mu_b)) \right]\ln\frac{\mu_b}{\mu_0} + \dots \;\;.
\end{eqnarray}
Here the number of flavors $n_f=5$, and $\mu_0=O(m_b)$ is an arbitrary normalization scale. The details of the calculation of Wilson coefficients is given in \cite{wilsoncoeff}. $F_\perp$, $G_\perp(x_{cb})$, $H_\perp(x_{cb})$ and $X_\perp$ are convolution integrals of hard-scattering kernels with the meson distribution amplitudes, and their values are summarized in Table \ref{tab:param}. In this table also appears the parameter $\displaystyle X=\ln(m_B/\Lambda_h)\,(1+\varrho\,e^{i\varphi})$, which parametrizes the logarithmically divergent integral $\int_0^1 dx/(1-x)$. $\varrho\le 1$ and the phase $\varphi$ are arbitrary. $\Lambda_h \approx 0.5$ GeV is a typical hadronic scale.\\
\\
\begin{table}[!t]
\begin{center}
\begin{tabular*}{160mm}{@{\extracolsep\fill}|c|c|c|c|c|}
\hline\hline
\multicolumn{5}{|c|}{CKM parameters and B meson mass}\\
\hline
$V_{us}$ & $V_{cb}$~~~ & $\left|V_{ub}/V_{cb}\right|$~~ & $\mbox{Re}(V_{us}^* V_{ub}/V_{cs}^*V_{cb})$~~~ & $m_B$~~~ \\
\hline
~~~~0.22~~~~ & $0.041 \pm 0.05~~~$ & $0.085 \pm 0.025$ & $0.011\pm 0.005~~~$ & 5.28 GeV~~~\\
\hline
\end{tabular*}
\\
\vspace*{0.1cm}
\begin{tabular*}{160mm}{@{\extracolsep\fill}|c|c|c|c|c|}
\hline
\multicolumn{5}{|c|}{$B$ meson parameters}\\
\hline
$f_B$ & $\lambda_B$~~ & $ a_{\perp} $~~ & $ < \bar v^{-1} >_\perp $~ & $h_{K^*}(x)~$\\
\hline
$200 \pm 20$ MeV & $350 \pm 150$ MeV~ & $ 0.19 \pm 0.02 $~~ & $ 3.7 \pm 0.04$~~ & $ ^{\displaystyle(4.8 \pm 0.5)~~}_{\displaystyle ~~+ (1.5 \pm 0.2)i} $ \\
\hline
\end{tabular*}
\\
\vspace*{0.1cm}
\begin{tabular*}{160mm}{@{\extracolsep\fill}|c|c|c|c|c|}
\hline
\multicolumn{5}{|c|}{$K^*$ meson parameters}\\
\hline
$F_{K^*}$ & $f_{K^*}^\perp$~~ & $m_{K^*}$~~ &  $f_{K^*}$~~ & $T_1^{B\to K^*}$~~ \\
\hline
~~$0.38 \pm 0.06$~~ & $175 \pm 9$ MeV~~ & 892 MeV~~ & $226 \pm 28$ MeV~~ & $0.30 \pm 0.05$~~\\
\hline
\end{tabular*}
\\
\vspace*{0.1cm}
\begin{tabular*}{160mm}{@{\extracolsep\fill}|c|c|c|c|}
\hline
\multicolumn{4}{|c|}{Convolution integral parameters}\\
\hline
$F_\perp$ & $G_\perp(x_{cb})$ & $H_\perp(x_{cb})$~~ & $X_\perp$~~\\
\hline
~~~~~$1.21\pm 0.06$~~~~~ & $^{\displaystyle(2.82\pm 0.20)~~~}_{\displaystyle~~~+(0.81\pm 0.23)i}$ & $^{\displaystyle(2.32\pm 0.16)~~~}_{\displaystyle~~~+(0.50\pm 0.18)i}$ & $^{\displaystyle(3.44\pm 0.47)\,X~~~}_{\displaystyle~~~-(3.91\pm 1.08)}$ \\
\hline
\end{tabular*}
\\
\vspace*{0.1cm}
\begin{tabular*}{160mm}{@{\extracolsep\fill}|c|c|c|c|c|}
\hline
\multicolumn{5}{|c|}{Quark and $W$-boson masses}\\
\hline
$m_b(m_b)$ & $m_c(m_b)$~ & $m_s$~~ & $m_t$~~ & $M_W$~\\
\hline
$4.2 \pm 0.1$ GeV & $1.2 \pm 0.2$ GeV & $0.10 \pm 0.03$ GeV & $172.5 \pm 2.7$ GeV & 80.4 GeV\\
\hline\hline
\end{tabular*}
\end{center}
\caption{The numerical values of the parameters.\label{tab:param}}
\end{table}%
The coefficient $a_7^c$ reads \cite{bosch}:
\begin{eqnarray}
a^c_7(K^*\gamma) &=& C_7(\mu_b) + \frac{\alpha_s(\mu_b) C_F}{4\pi} \left( C_1(\mu_b) G_1(x_{cb})+ C_8(\mu_b) G_8\right)\\
&& +\frac{\alpha_s(\mu_h) C_F}{4\pi} \left( C_1(\mu_h) H_1(x_{cb})+ C_8(\mu_h) H_8\right) \nonumber\;\;,
\end{eqnarray}
where $\mu_h=\sqrt{\Lambda_h \mu_b}$ is the spectator scale, and
\begin{equation}
G_1(x_{cb}) = -\frac{104}{27}\ln\frac{\mu_b}{m_b}+ g_1(x_{cb}) \;\; , \;\;G_8 = \frac{8}{3}\ln\frac{\mu_b}{m_b} + g_8 \;\;.
\end{equation}
We have
\begin{eqnarray}
g_8 &=& \frac{11}{3}-\frac{2\pi^2}{9}+\frac{2i\pi}{3}\;\;,\\
g_1(x) &=& \frac{2}{9} \bigg[ 48+30i\pi-5\pi^2-2i\pi^3 -36\zeta_3 +\left( 36+6i\pi-9\pi^2\right)\ln x\nonumber\\
&&+\left( 3+6i\pi\right) \ln^2\! x+\ln^3\! x \bigg] x\\
&& {}+\frac{2}{9} \bigg[ 18+2\pi^2 -2i\pi^3 +\left( 12-6\pi^2 \right)\ln x +6i\pi\ln^2\! x+\ln^3\! x\bigg] x^2\nonumber \\
&& {}+\frac{1}{27} \bigg[ -9+112 i\pi-14\pi^2+\left(182-48i\pi\right)\ln x-126\ln^2\! x\bigg] x^3 \nonumber\\
&&-\frac{833}{162}-\frac{20i\pi}{27} +\frac{8\pi^2}{9} x^{3/2}\;\;,  \nonumber
\end{eqnarray}
where $\zeta_3 \approx 1.2020569$ and $x_{cb}=\displaystyle\frac{m^2_c}{m^2_b}\;$. We also need:
\begin{equation}
H_1(x)=-\frac{2\pi^2}{3 N}\frac{f_B f^\perp_{K^*}}{F_{K^*} m^2_B}\int^1_0 d\xi\frac{\Phi_{B1}(\xi)}{\xi}\int^1_0 dv\, h(\bar v,x)\Phi_\perp(v) \;\;,
\end{equation}
where $h(u,x)$ is the hard-scattering function:
\begin{equation}
h(u,x)=\displaystyle\frac{4x}{u^2}\left[\mbox{Li}_2\!\left(\displaystyle\frac{2}{1-\sqrt{\displaystyle\frac{u-4x+i\varepsilon}{u}}}\right)+\mbox{Li}_2\!\left(\displaystyle\frac{2}{1+\sqrt{\displaystyle\frac{u-4x+i\varepsilon}{u}}}\right)\right]-\frac{2}{u} \;\;,
\end{equation}
and where $\mbox{Li}_2$ is the usual dilogarithm function $\displaystyle\mbox{Li}_2(x)=-\int^x_0 dt\frac{\ln(1-t)}{t} \;$.\\
$\Phi_\perp$ is the light-cone wave function with transverse polarization and $\Phi_{B1}$ is a distribution amplitude of the $B$ meson involved in the leading-twist projection. We have also:
\begin{equation}
H_8=\frac{4\pi^2}{3 N}\frac{f_B f^\perp_{K^*}}{F_{K^*} m^2_B}\int^1_0 d\xi\frac{\Phi_{B1}(\xi)}{\xi}\int^1_0 dv\frac{\Phi_\perp(v)}{v} \;\;.
\end{equation}
The first negative moment of $\Phi_{B1}$ can be parametrized by a quantity $\lambda_B$ such as 
\begin{equation}
\int^1_0 d\xi\frac{\Phi_{B1}(\xi)}{\xi}=\frac{m_B}{\lambda_B}\;\,. 
\end{equation}
Following \cite{parkho}, and defining
\begin{equation}
\left < f \right >_{\perp} \equiv \int\limits_0^1 dv \, f (v) \, \phi_{\perp} (v) \;\;, 
\end{equation}
we can write
\begin{equation}
\int^1_0 dv\, h(\bar v,x)\Phi_\perp(v) = -2 \left( < \bar v^{-1} >_\perp - h_{K^*}(x) \right) \;\;,
\end{equation}
where $\bar v = 1 -v$, and
\begin{equation}
\int^1_0 dv\frac{\Phi_\perp(v)}{v} = - 6\, a_{\perp} + < \bar v^{-1} >_\perp \;\;.
\end{equation}
The numerical values of all the needed parameters can be found in Table~\ref{tab:param}.\\
\\
Using the equations mentioned in this section, \verb?SuperIso? is able to calculate the isospin asymmetry of $B \rightarrow K^* \gamma$. For the computation of the inclusive branching ratio, we used the relations of \cite{wilsoncoeff,bsg}.
%
\section{Content of the SuperIso package}%
\label{content}
\verb?SuperIso? is a C program respecting the C99 standard (through the use of the \verb?complex.h? library), whose main purpose is to calculate the isospin symmetry breaking of $B \rightarrow K^* \gamma$ decays in the MSSM with minimal flavor violation. In this package, five main programs are provided, but the users are also invited to write their own main programs. \verb?main_example.c? is intended as an example for writing new programs, and \verb?slha.c? can scan files written following the SUSY Les Houches Accord (SLHA) \cite{slha}, and calculates the corresponding isospin asymmetry. The main programs \verb?msugra.c?, \verb?amsb.c? and \verb?gmsb.c? have to be linked to the \verb?ISASUGRA/ISAJET? \cite{isajet} and/or the \verb?SOFTSUSY? \cite{softsusy} packages, in order to compute mass spectra and couplings within the minimal Supergravity scenario (mSUGRA), the minimal Anomaly Mediated SUSY Breaking scenario (AMSB) or the minimal Gauge Mediated SUSY Breaking scenario (GMSB).\\
The calculation of isospin asymmetry requires the following steps:
\begin{itemize}
 \item Generation of a SLHA file with \verb?ISAJET? or \verb?SOFTSUSY?, or the SLHA file provided by the user,
 \item Scan of the SLHA file,
 \item Computation of the required Wilson coefficients,
 \item Calculation of the isospin asymmetry.
\end{itemize}
\verb?SuperIso? also computes the inclusive branching ratio of $b \rightarrow s \gamma$, and the code is modular enough so that routines calculating other observables, such as the branching ratio of $B_s \rightarrow \mu^+ \mu^-$ or the anomalous magnetic moment of the muon, can be easily added. For this purpose, a procedure has to be added in \verb?src/? which can also benefit from the functions already implemented in the program (for example for the calculation of the Wilson coefficients and the scan of the SLHA files.)\\
\\
The code includes the definition of a structure in \verb?src/include.h?:
\begin{verbatim}
typedef struct parameters
{
	int model; /* mSUGRA = 1, GMSB = 2, AMSB = 3 */
	int generator; /* ISAJET = 1, SOFTSUSY = 2 */
	float Q; /* Qmax ; default = M_EWSB = sqrt(m_stop1*mstop2) */
	float m0,m1_2,tan_beta,sign_mu,A0,mass_W; /* mSUGRA parameters */
	float Lambda,Mmess,N5,cgrav,m3_2; /* AMSB, GMSB parameters */
	float mass_Z,mass_b,mass_top_pole,mass_tau_pole; /* SM parameters */
	float inv_alpha_em,alpha_s_MZ,alpha,Gfermi,GAUGE_Q; /* SM parameters */
	float charg_Umix[3][3],charg_Vmix[3][3],stop_mix[3][3],sbot_mix[3][3]
,stau_mix[3][3],neut_mix[5][5],mass_neut[5]; /* mass mixing matrices */
	float Min,M1_Min,M2_Min,M3_Min,At_Min,Ab_Min,Atau_Min,M2H1_Min,M2H2_Min,
mu_Min,M2A_Min,tb_Min,mA_Min; /* optional input parameters at scale Min */
	float MeL_Min,MmuL_Min,MtauL_Min,MeR_Min,MmuR_Min,
MtauR_Min; /* optional input parameters at scale Min */
	float MqL1_Min,MqL2_Min,MqL3_Min,MuR_Min,McR_Min,MtR_Min,MdR_Min,MsR_Min,
MbR_Min; /* optional input parameters at scale Min */
	float N51,N52,N53,M2H1_Q,M2H2_Q; /* optional input 
parameters (N51...3: GMSB) */
	float mass_d,mass_u,mass_s,mass_c,mass_t,mass_e,mass_nue,mass_mu,mass_num,
mass_tau,mass_nut; /* SM masses */
	float mass_gluon,mass_photon,mass_Z0; /* SM masses */
	float mass_h0,mass_cH0,mass_A0,mass_H,mass_dnl,mass_upl,mass_stl,mass_chl,
mass_b1,mass_t1; /* Higgs & superparticle masses */
	float mass_el,mass_nuel,mass_mul,mass_numl,mass_tau1,mass_nutl,mass_gluino,
mass_cha1,mass_cha2; /* Higgs & superparticle masses */
	float mass_dnr,mass_upr,mass_str,mass_chr,mass_b2,mass_t2,mass_er,mass_mur,
mass_tau2; /* superparticle masses */
	float mass_nuer,mass_numr,mass_nutr,mass_graviton,
mass_gravitino; /* superparticle masses */
	float gp,g2,g3,YU_Q,yut,YD_Q,yub,YE_Q,yutau; /* Yukawa couplings */
	float HMIX_Q,mu_Q,tanb_GUT,Higgs_VEV,mA2_Q,MSOFT_Q,M1_Q,M2_Q,
M3_Q; /* parameters at scale Q */
	float MeL_Q,MmuL_Q,MtauL_Q,MeR_Q,MmuR_Q,MtauR_Q,MqL1_Q,MqL2_Q,
MqL3_Q,MuR_Q,McR_Q,MtR_Q,MdR_Q,MsR_Q,MbR_Q; /* masses at scale Q */
	float AU_Q,A_u,A_c,A_t,AD_Q,A_d,A_s,A_b,AE_Q,A_e,A_mu,
A_tau;  /* trilinear couplings */
}
parameters;
\end{verbatim}
This structure contains all the important parameters and is called by several functions in the program. We can now review the important routines of the code.
\begin{itemize}
 \item \verb?void Init_param(struct parameters* param)?\\
\\
This function initializes the \verb?param? structure, setting the parameters at 0, apart for the SM masses and the value of the strong coupling constant at the $Z$-boson mass, which receive the values given in \cite{PDG}. Moreover, \verb?model? is set to -1 as an indicator for a wrong reading of the SLHA file. After the reading step, a value remaining -1 indicates a problem with the SLHA file.\\

 \item \verb?int Les_Houches_Reader(char name[], struct parameters* param)?\\
\\
This routine is able to read a SLHA file whose name is contained in \verb?name?, and put all the read parameters in the structure \verb?param?. This function has been written based on \cite{slha}. It can be reused independently from \verb?SuperIso?, provided \verb?struct parameters? is defined.\\

 \item \verb?int softsusy_sugra(float m0, float m12, float tanb, float A0,?\\
\verb?float sgnmu, float mtop, float mbot, float alphas_mz, char name[])?
 \item \verb?int isajet_sugra(float m0, float m12, float tanb, float A0, ?\\
\verb?float sgnmu, float mtop, char name[])?
 \item \verb?int softsusy_gmsb(float Lambda, float Mmess, float tanb, int N5,?\\
\verb? float cGrav, float sgnmu, float mtop, float mbot, float alphas_mz,?\\
\verb? char name[])?
 \item \verb?int softsusy_amsb(float m0, float m32, float tanb, float sgnmu,?\\
\verb?float mtop, float mbot, float alphas_mz, char name[])?\\
\\
The above four routines call \verb?SOFTSUSY? or \verb?ISAJET? to compute the mass spectrum corresponding to the input parameters. The input parameters are described in the next section. These routines return a SLHA file whose name has to be specified in the string \verb?name?.\\

\item \verb?float alpha_s_running(float Q, float mtop, float mbot,?\\
\verb?float alphas_MZ, float MZ)?
\\
This function computes the strong coupling constant at the energy scale \verb?Q?, provided the top quark mass \verb?mtop?, bottom quark mass \verb?mbot?, $Z$-boson mass ($M_Z$) \verb?MZ? and $\alpha_s(M_Z)$ \verb?alphas_MZ? are given, so that the matching can be made between the scales corresponding to different flavor numbers. The main formula for calculating $\alpha_s$ is based on \cite{PDG}.\\

\item \verb?float running_mass(float quark_mass, float Qinit, float Qfin,?\\
\verb?float mtop, float mbot, float alphas_MZ, float MZ)?\\
\\
This function calculates the running quark mass at the scale \verb?Qfin?, for a quark of mass \verb?quark_mass? at the scale \verb?Qinit?, knowing the top quark mass \verb?mtop?, bottom quark mass \verb?mbot?, $Z$-boson mass ($M_Z$) \verb?MZ? and $\alpha_s(M_Z)$ \verb?alphas_MZ?, based on \cite{running}.\\

\item \verb?void C_calculator(float C0[], float C1[],float mu,?\\
\verb?struct parameters* param)?\\
\\
This procedure computes the LO contributions to the Wilson coefficients $C_{1 \cdots 8}$ \verb?C0[]? as well as the NLO contributions \verb?C1[]? at the energy scale \verb?mu?, using the parameters of \verb?param?, and based on \cite{wilsoncoeff}.\\

\item \verb?float delta0m(float C0[],float C0_spec[],float C1[],?\\
\verb?float C1_spec[],struct parameters* param,float mub,float muspec,?\\
\verb?float lambda_h)?\\
\\
This function computes the isospin asymmetry as described in the precedent section, using both the LO and NLO parts of the Wilson coefficients at scale \verb?mub? (\verb?C0[]? and \verb?C1[]?), and at the spectator scale \verb?muspec? (\verb?C0_spec[]? and \verb?C1_spec[]?), with the additional input $\Lambda_h$ \verb?lambda_h?, in GeV. It uses the \verb?complex.h? library which is available in C compilers respecting the C99 standard.\\
\\
\item \verb?float delta0_calculator(char name[])?\\
\\
This function is somehow a ``container'' function whose argument is the name of the SLHA file, and which calls first \verb?Init_param? and \verb?Les_Houches_Reader?, then calculates the Wilson coefficients with \verb?C_calculator?, and finally returns the results of \verb?delta0m?. \verb?main_example.c? provides an example of computing the isospin asymmetry without calling this function, but by calling \verb?delta0m? directly. This is useful if the user wants to provide his/her own Wilson coefficients as input.\\
\\
\item \verb?float BRbsgamma_calculator(char name[])?\\
\\
This function, similar to the precedent one, computes the inclusive branching ratio of $b \rightarrow s \gamma$ (based on \cite{bsg}). It has to call two additional functions, \verb?Cem_calculator? and \verb?BR_bsgamma?.\\
\\
\item \verb?int charged_LSP_calculator(char name[])?\\
This routine, whose argument is the name of the SLHA file, checks whether the lightest supersymmetric particle (LSP) is charged, and if so returns 0, and if not returns 1 (getting -1 means that a problem occurred). Note that if gravitino is the LSP, then this routine verifies the charge of the NLSP.\\
\\
\item \verb?int excluded_mass_calculator(char name[])?\\%
\begin{table}
\begin{center}
\begin{tabular}{|c|c|c|c|c|c|c|c|c|c|c|} \hline
Particle & $h^0$ & $\chi^0_1$ & $\tilde{l}_R$ & $\tilde{\nu}_{e,\mu}$ & $\chi^\pm_1$ & $\tilde{t}_1$ & $\tilde{g}$ & $\tilde{b}_1$ & $\tilde{\tau}_1$ & $\tilde{q}_R$\\
\hline
Lower bound & 111 & 46 & 88 & 43.7 & 67.7 & 92.6 & 195 & 89 & 81.9 & 250\\
\hline
\end{tabular}
\end{center}
\caption{Lower bounds on the particle masses in GeV \cite{PDG}.}
\label{tab:masses}
\end{table}%
This routine, whose argument is the name of the SLHA file, checks whether the parameter space point is excluded by the collider constraints on the particle masses summarized in table~\ref{tab:masses}, and if so returns 1, otherwise returns 0 (getting -1 means that a problem occurred). The constraints in table~\ref{tab:masses} can be easily updated by modifying \verb?src/masses.c?.
\end{itemize}
%
\section{Compilation and installation instructions}%
\label{compilation}
The \verb?SuperIso? package can be downloaded at\\
{\tt http://www3.tsl.uu.se/$\sim$nazila/superiso/}\\
After unpacking the package, the main directory,\\
\verb?superiso_vX.X?\\
is created. It contains the \verb?src/? directory, in which all the source files can be found. The main directory contains also a \verb?Makefile?, a \verb?README?, five sample main programs (\verb?main_example.c?, \verb?msugra.c?, \verb?amsb.c?, \verb?gmsb.c? and \verb?slha.c?) and one example of input file in the SUSY Les Houches Accord format (\verb?example.lha?). The compiler options should be defined in the \verb?Makefile?, as well as the path to the ISAJET \verb?isasugra.x? and SOFTSUSY \verb?softpoint.x? executable files, when needed. This package has been written for a C compiler respecting the C99 standard. In particular, it has been tested successfully with the GNU C Compiler and the Intel C Compiler on Linux 32-bits or 64-bits machines, and with \verb?SOFTSUSY? 2.0.14 and \verb?ISAJET? 7.75. Additional information can be found in the \verb?README? file.\\
To compile the library, type\\
\verb?make?\\
\noindent This creates the \verb?libisospin.a? in \verb?src/?. Then, to compile one of the five programs provided in the main directory, type:
\verb?make name?\\
\noindent where \verb?name? can be \verb?main_example?, \verb?msugra?, \verb?amsb?, \verb?gmsb? or \verb?slha?. This creates an executable with the \verb?.x? extension. Note that \verb?slha? does not need \verb?ISAJET? or \verb?SOFTSUSY? programs, but \verb?msugra?, \verb?amsb? and \verb?gmsb? need at least one of them. Beside, \verb?main_example? is somehow a test--program in which \verb?ISAJET? and \verb?SOFTSUSY? have been disabled by default, but their use can be restored by uncommenting 
\verb?#define USE_ISAJET?
\noindent and
\verb?#define USE_SOFTSUSY?
\noindent at the beginning of the program.\\
\\
\verb?main_example.x? has to be run without any argument, and performs the calculation of test--points using the main routines of the library. It can be used as a basis for writting the user's own main program.\\
\\
\verb?slha.x? calculates the isospin asymmetry and the inclusive branching ratio using the parameters contained in the SLHA file whose name has to be passed as input parameter.\\
\\
\verb?amsb.x?, \verb?gmsb.x? and \verb?msugra.x? also calculate the isospin asymmetry and the inclusive branching ratio, starting first by calculating the mass spectrum and couplings thanks to \verb?ISAJET? (for \verb?msugra.x? only) and/or \verb?SOFTSUSY? within respectively the AMSB, GMSB and mSUGRA parameter spaces. For all these programs, arguments referring to the usual input parameters have to be passed to the program. If not, a message will describe which parameters have to be specified.\\
\\
A more detailled presentation of the input and output of these modules is given in the next section.
%
\section{Sample input and output}%
\label{sample}
The inputs and outputs of the five provided main programs are described in the following. However, a 0-value for the calculation of isospin asymmetry or branching ratio corresponds to a problem in reading the SLHA file, which either has not been generated by \verb?ISAJET? or \verb?SOFTSUSY? for any reason, or is not conformed to the SLHA format.
\begin{itemize}
 \item \verb?main_example.x?\\
This program is an example demonstrating the capabilities of \verb?SuperIso?. No input argument is needed as the calls to the procedure are hardcoded. If\\
\verb?#define USE_ISAJET?\\
\verb?#define USE_SOFTSUSY?\\
is uncommented at the beginning of the program, it will call first the functions \verb?isajet_sugra?, \verb?softsusy_sugra?, \verb?softsusy_amsb? and \verb?softsusy_gmsb?, with an adequate set of input parameters and calculate the isospin asymmetry and branching ratio thanks to the routines \verb?delta0_calculator? and \verb?BRbsgamma_calculator?. It will then calculate the isospin asymmetry and branching ratio corresponding to the parameters of the \verb?example.lha? file. Finally, input parameters ({\it i.e.} the Wilson coefficients, the masses of strange, charm, bottom and top quarks, the $Z$-boson mass $M_Z$ and $\alpha_s(M_Z)$) corresponding to the Standard Model case are given directly in the program, and the isospin asymmetry is calculated using the \verb?delta0m? function. This last calculation is  provided as an example of calculations of isospin asymmetry without a SLHA file or without \verb?ISAJET? or \verb?SOFTSUSY?. If the \verb?#define?'s are uncommented, the output is for example\\
\begin{verbatim}
delta0_isajet_sugra=0.101417
BR_isajet_sugra=0.000257

delta0_softsusy_sugra=0.100604
BR_softsusy_sugra=0.000260

delta0_softsusy_amsb=0.096156
BR_softsusy_amsb=0.000286

delta0_softsusy_gmsb=0.079418
BR_softsusy_gmsb=0.000393

delta0_example_slha=0.100568
BR_example_slha=0.000261

delta0_direct_SM=0.086136

\end{verbatim}
\newpage
 \item \verb?slha.x?\\
This program calculates the isospin asymmetry and inclusive branching ratio while reading the needed parameters in SLHA files. It also checks whether the SUSY parameter space point is excluded by the experimental mass limits. For example, the command\\
\begin{verbatim}
./slha.x example.lha
 
\end{verbatim}
returns\\
\begin{verbatim}
delta0=0.100568
BR=0.000261
excluded_mass=0

\end{verbatim}

\item \verb?msugra.x?\\
This program generates the isospin asymmetry and inclusive branching ratio corresponding to the mSUGRA parameters generated by \verb?ISAJET? and/or \verb?SOFTSUSY?. It also checks whether the SUSY parameter space point is excluded by the experimental mass limits, or whether the LSP is charged. If only one of these generators is available the corresponding \verb?#define? has to be commented. The necessary arguments to this program are:\\
\begin{itemize}
 \item $m_0$: universal scalar mass at GUT scale,
 \item $m_{1/2}$: universal gaugino mass at GUT scale,
 \item $A_0$: trilinear soft breaking parameter at GUT scale,
 \item $\tan\beta$: ratio of the two Higgs vacuum expectation values.\\
\end{itemize}
Optional arguments can also be given:\\
\begin{itemize}
 \item $sign(\mu)$: sign of Higgsino mass term, positive by default
 \item $m_t^{pole}$: top quark pole mass, by default 172.5 GeV,
 \item $\overline{m_b}(\overline{m_b})$: scale independent b-quark mass, by default 4.2 GeV (option only available for \verb?SOFTSUSY?),
 \item $\alpha_s(M_Z)$: strong coupling constant at the $Z$-boson mass, by default 0.1172 (option only available for \verb?SOFTSUSY?).\\
\end{itemize}
With \verb?SOFTSUSY? 2.0.14 and \verb?ISAJET? 7.75, running the program with:\\
\begin{verbatim}
./msugra.x 500 500 0 50

\end{verbatim}
returns\\
\begin{verbatim}
delta0_softsusy=0.100604
BR_softsusy=0.000260
charged_LSP_softsusy=0
excluded_masses_softsusy=0

delta0_isajet=0.101457
BR_isajet=0.000257
charged_LSP_isajet=0
excluded_masses_isajet=0

\end{verbatim}

\item \verb?amsb.x?\\
This program generates the isospin asymmetry and inclusive branching ratio corresponding to the AMSB parameters generated by \verb?SOFTSUSY?. It also checks whether the SUSY parameter space point is excluded by the experimental mass limits. The necessary arguments to this program are:\\
\begin{itemize}
 \item $m_0$: universal scalar mass at GUT scale,
 \item $m_{3/2}$: gravitino mass at GUT scale,
 \item $\tan\beta$: ratio of the two Higgs vacuum expectation values.\\
\end{itemize}
Optional arguments are the same as for the mSUGRA case.
With \verb?SOFTSUSY? 2.0.14, running the program with:\\
\begin{verbatim}
./amsb.x 400 10000 30 -1

\end{verbatim}
returns\\
\begin{verbatim}
delta0=0.110040
BR=0.000224
excluded_mass=1

\end{verbatim}

\item \verb?gmsb.x?\\
This program generates the isospin asymmetry and inclusive branching ratio corresponding to the GMSB parameters generated by \verb?SOFTSUSY?. The necessary arguments to this program are:\\
\begin{itemize}
 \item $\Lambda$: scale of the SUSY breaking in GeV (usually 10,000 -- 100,000 GeV),
 \item $M_{mess}$: messenger mass scale ($> \Lambda$),
 \item $N_5$: equivalent number of $5+\bar{5}$ messenger fields,
 \item $\tan\beta$: ratio of the two Higgs vacuum expectation values.\\
\end{itemize}
Optional arguments are the same as for the mSUGRA case, with an additional one:\\
\begin{itemize}
 \item $c_{Grav}$ ($\ge 1$): ratio of the gravitino mass, to its value for a breaking scale of $\Lambda$, 1 by default.\\
\end{itemize}
With \verb?SOFTSUSY? 2.0.14, running the program with:\\
\begin{verbatim}
./gmsb.x 60000 80000 5 50

\end{verbatim}
returns\\
\begin{verbatim}
delta0=0.079901
BR=0.000390
excluded_mass=0

\end{verbatim}
\end{itemize}
%
\section{Results}
\label{result}
\begin{figure}[pt]
\begin{center}
\includegraphics[width=10cm]{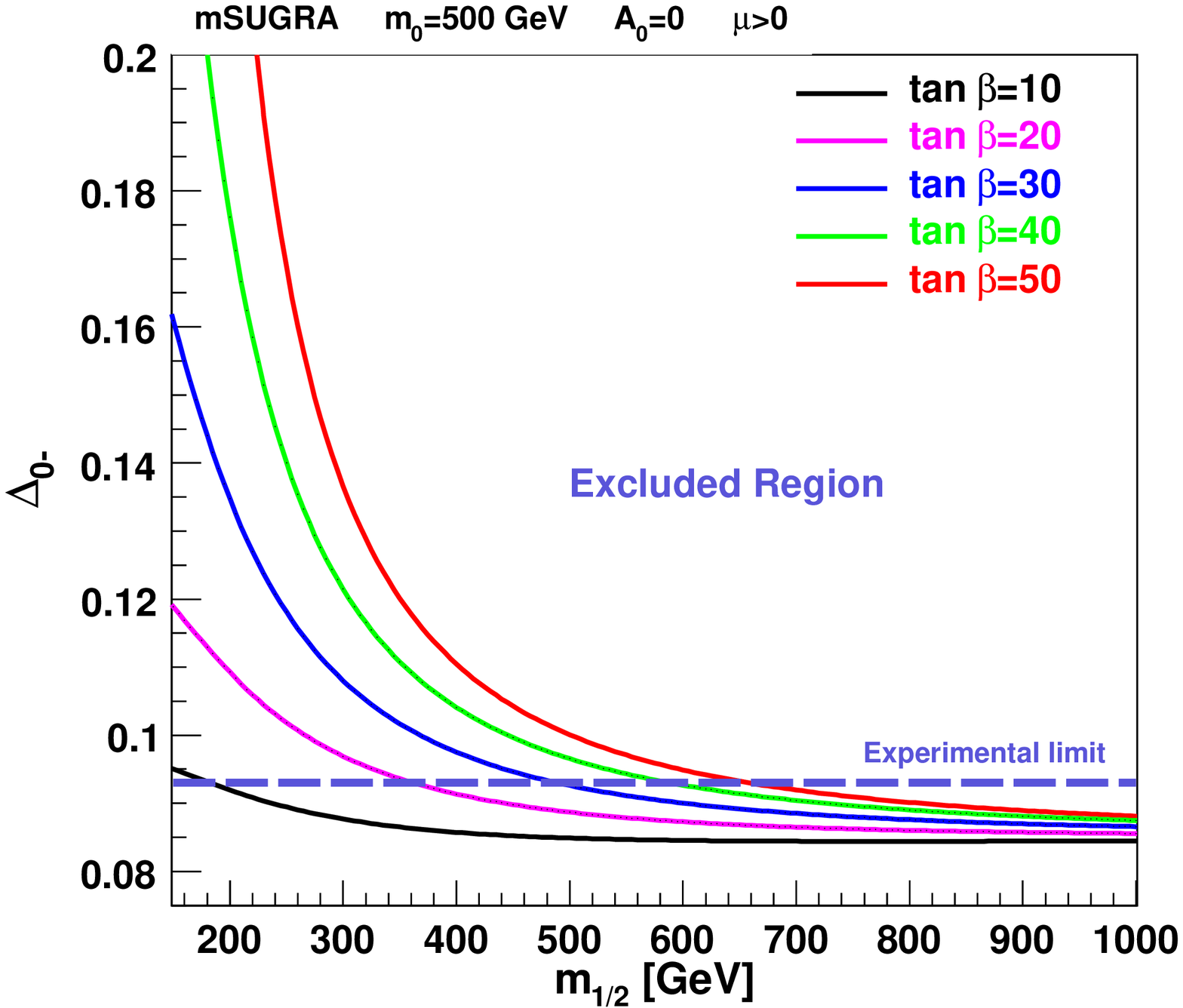}\\
\vspace*{1.0cm}
\includegraphics[width=10cm]{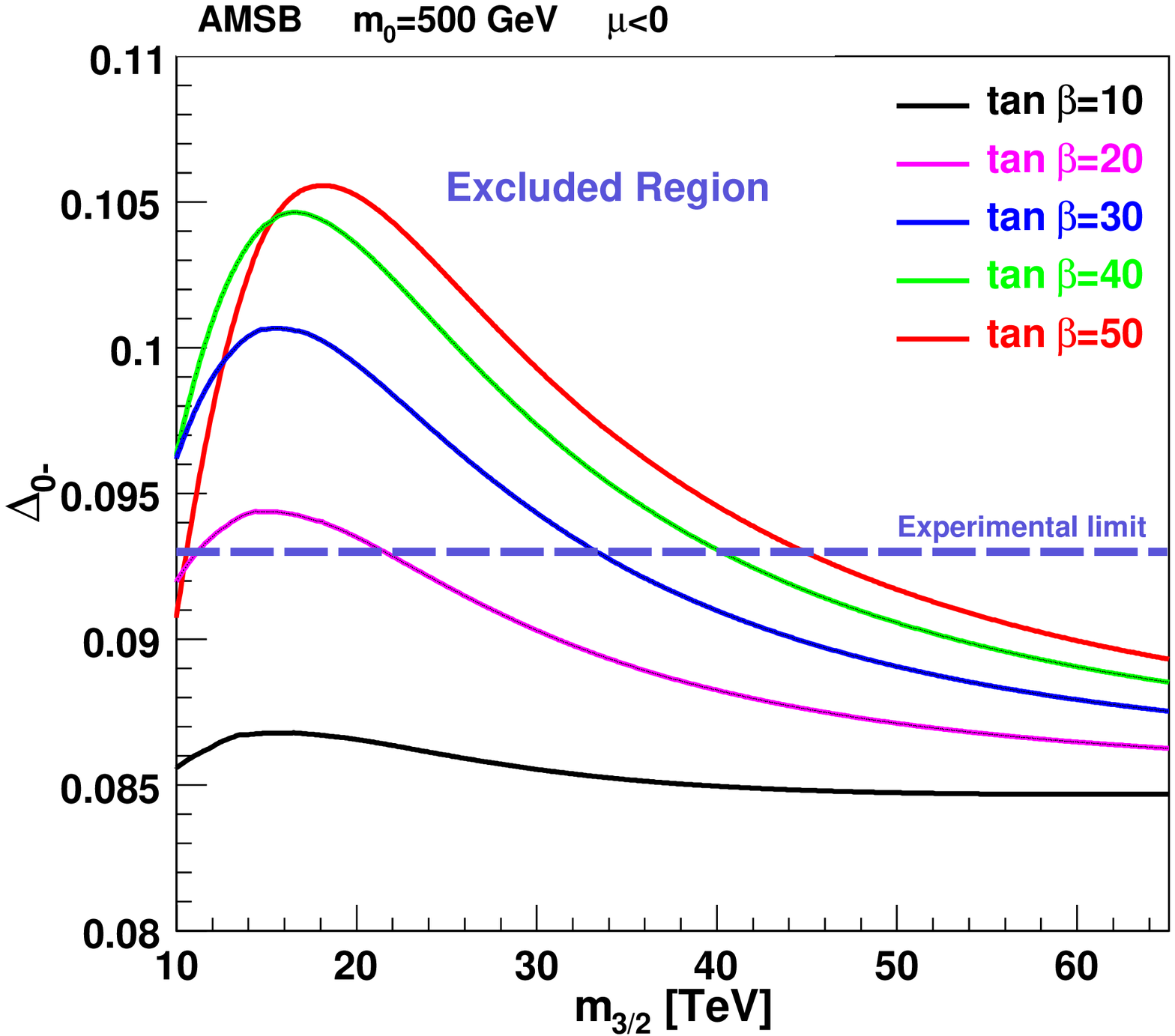}
\caption{Isospin asymmetry in function of $m_{1/2}$ in the mSUGRA parameter space (in the top) and of $m_{3/2}$ in the AMSB parameter space (in the bottom), for several values of $\tan\beta$. The dashed violet horizontal line corresponds to the experimental limit of \cite{belle,babar}.}
\label{fig}
\end{center}
\end{figure}

The \verb?SuperIso? code has been extensively tested, and compared to the calculation of the isospin asymmetry available in the literature for the Standard Model \cite{kagan}, and with the Wilson coefficients and inclusive branching ratios calculated for example by micrOMEGAS \cite{micromegas}, and they globally show a good agreement.\\
\\
Figure~\ref{fig} shows the dependence of the isospin asymmetry in function of the parameters of the mSUGRA and AMSB parameter spaces, and reveals that isospin symmetry breaking can strongly constrain the supersymmetric parameters space, especially for high $\tan \beta$ value.\\
A more complete study of the constraints by isospin asymmetry in different supersymmetric parameter spaces is presented in \cite{mahmoudi}.
%
\section{Conclusion}
The \verb?SuperIso? package allows to calculate the isospin asymmetry of $B \rightarrow K^* \gamma$ in the MSSM with minimal flavor violation, for any file in the SLHA format. It can also call \verb?ISAJET? or \verb?SOFTSUSY? to generate a SLHA parameter file in the mSUGRA, AMSB or GMSB supersymmetric parameter space. The package also provides a routine to calculate the inclusive branching ratio of $B \rightarrow X_s \gamma$ to enable a comparison between the constraints on the supersymmetric parameters from the branching ratio and from the isospin asymmetry.\\
%
\section*{Acknowledgments}
\noindent The \verb?SuperIso? program has largely benefited from the help of Alexandre Arbey. Thanks also to the members of the THEP group at Uppsala University for the useful discussions.

\end{document}